\documentclass[prb,twoside,twocolumn,showpacs,superscriptaddress]{revtex4}
\usepackage{graphicx}
\usepackage{amsmath}
\usepackage{amssymb}

\begin{document}

\title{Anisotropy and internal field distribution of MgB$_2$ in the mixed state\\ at low temperatures}
\author{M. Angst}
  \email[Email: ]{angst@ameslab.gov}
\affiliation{Physik-Institut der Universit\"at Z\"urich, CH-8057
Z\"urich, Switzerland}
\affiliation{Ames Laboratory and Department of Physics and
Astronomy, Iowa State University,Ames,IA 50011,USA}

\author{D. Di Castro}
\affiliation{Physik-Institut der Universit\"at Z\"urich, CH-8057
Z\"urich, Switzerland}

\author{D.~G. Eshchenko}
\affiliation{Physik-Institut der Universit\"at Z\"urich, CH-8057
Z\"urich, Switzerland}

\affiliation{Paul Scherrer Institute, CH-5232 Villigen PSI,
Switzerland}

\author{R. Khasanov}
\affiliation{Paul Scherrer Institute, CH-5232 Villigen PSI,
Switzerland}

\affiliation{DPMC, Universit\'{e} de Gen\`{e}ve, 24 Quai
Ernest-Ansermet, CH-1211 Gen\`{e}ve, Switzerland}

\affiliation{Physik-Institut der Universit\"at Z\"urich, CH-8057
Z\"urich, Switzerland}

\author{S.~Kohout}
\affiliation{Physik-Institut der Universit\"at Z\"urich, CH-8057
Z\"urich, Switzerland}

\author{I.~M. Savic}
\affiliation{Faculty of Physics, University of Belgrade, 11001
Belgrade, Yugoslavia}

\author{A. Shengelaya}
\affiliation{Physik-Institut der Universit\"at Z\"urich, CH-8057
Z\"urich, Switzerland}

\author{S.~L. Bud'ko}
\author{P.~C. Canfield}
\affiliation{Ames Laboratory and Department of Physics and
Astronomy, Iowa State University,Ames,IA 50011,USA}

\author{J.~Jun}
\author{J.~Karpinski}
\author{S.~M. Kazakov}
\affiliation{Solid State Physics Laboratory, ETH, CH-8093
Z\"urich, Switzerland}

\author{R.~A. Ribeiro}
\affiliation{Ames Laboratory and Department of Physics and
Astronomy, Iowa State University,Ames,IA 50011,USA}

\author{H.~Keller}
\affiliation{Physik-Institut der Universit\"at Z\"urich, CH-8057
Z\"urich, Switzerland}

\date{\today}
\begin{abstract}
Magnetization and muon spin relaxation on MgB$_2$ were measured as
a function of field at $2\,{\rm K}$. Both indicate an
inverse-squared penetration depth strongly decreasing with
increasing field $H$ below about $1\,{\rm T}$. Magnetization also
suggests the anisotropy of the penetration depth to increase with
increasing $H$, interpolating between a low $H_{c1}$ and a high
$H_{c2}$ anisotropy. Torque vs angle measurements are in agreement
with this finding, while also ruling out drastic differences
between the mixed state anisotropies of the two basic length
scales penetration depth and coherence length.
\end{abstract}
\pacs{74.25.Op, 74.20.De, 74.25.Ha, 74.70.Ad}
\maketitle

\label{intro} The understanding of the physical properties of the
recently discovered $39 \, {\text{K}}$
superconductor\cite{Nagamatsu01} MgB$_2$ has made rapid progress
in the last 3 years.\cite{Canfield02rev} A central issue of
research has been the involvement in superconductivity of two sets
of bands with different dimensionality and pairing
strength.\cite{Liu01,Bouquet01b,Gonnelli02} This ``two-band
superconductivity'' leads to an array of unusual superconducting
properties such as specific heat,\cite{Bouquet01b} particularly
also to a very unusual behavior of the superconducting
anisotropies.\cite{Angst03Nova}

For example, a pronounced temperature $T$ dependence of the
anisotropy $\gamma_H$ of the upper critical field $H_{c2}$,
directly related to the coherence length $\xi$, was
observed\cite{MgB2anisPRL02,Sologubenko02} and calculated based on
the two-band model.\cite{Miranovic03,Golubov03} Strikingly,
calculations of the low field penetration depth anisotropy
$\gamma_{\lambda}$, predicted a much lower anisotropy of this
quantity, with a $T$ dependence opposite to the one of
$H_{c2}$.\cite{Kogan02} This was experimentally confirmed as well,
based on measurements, e.g., of $H_{c1}$,\cite{Lyard04,Kim04} by
small angle neutron scattering (SANS)\cite{Cubitt03,Cubitt03b} and
scanning tunneling spectroscopy (STS).\cite{Eskildsen03} However,
e.g.\ the experiment of Ref.\ \onlinecite{Cubitt03b} indicates
that whereas in the limit of very low fields $H$
$\gamma_{\lambda}$ is indeed close to $1$, it is rising with
increasing $H$, as deduced earlier more
indirectly.\cite{MgB2anisPRL02}

The behavior of the anisotropies of the length scales in the mixed
state $H_{c1} \! < \! H \! < \! H_{c2}$ still needs to be
clarified. One point of view \cite{Kogan02b} surmises constant
(with respect to $H$) anisotropies of the penetration depth
$\gamma_{\lambda}$ and the coherence length $\gamma_H$, which are,
however, different from each other. This difference was predicted
to lead to a sign reversal in the angle dependent
torque.\cite{Kogan02b} Another point of view is that these
anisotropies are not drastically different from each other, but
both increase with increasing field, interpolating from the
$H_{c1}$ anisotropy in low fields to the $H_{c2}$ anisotropy in
high fields.\cite{note_twobandanis}

Here, we support the latter point of view by analyzing SQUID
(superconducting quantum interference device) and torque
magnetization data measured on a MgB$_2$ single crystal with very
low pinning, and muon spin relaxation ($\mu$SR) data measured on
randomly aligned MgB$_2$ powder. In the absence of a more
elaborate model of the mixed state of a two-band
superconductor,\cite{note_twobandanis} we base the analysis on the
London model, allowing however for a $H$ dependent penetration
depth, which is obtained from the $H$ or angle $\theta$ dependence
of the bulk magnetization (SQUID/torque) as well as from the
average variation of the internal field ($\mu$SR). From SQUID and
$\mu$SR we find a rapid decrease of the inverse-squared
penetration depth $1/\lambda ^2$ (``superfluid density'') with
$\mu_{\circ} H$ increasing below about $1\,{\rm T}$, and SQUID and
torque data agree on the anisotropy $\gamma_{\lambda}$ increasing
strongly with $H$. The analysis of the torque data further
suggests that $\gamma_H$ is not very different from
$\gamma_{\lambda}$.

\label{exp} Single crystals of MgB$_2$ were grown with a high
pressure cubic anvil technique,\cite{Karpinski03SST}
and a crystal with particularly low pinning was selected for
measurements with a Quantum Design MPMS-XL SQUID magnetometer and
a noncommercial torque magnetometer.\cite{Willemin98b} The crystal
has a wedge shape, with one of the faces parallel to the $ab$
planes. The $\mu$SR experiment on poly-crystalline MgB$_2$ was
similar to the one of Ref.\ \onlinecite{DiCastro03}.

\label{res}

\begin{figure}[tb]
\includegraphics[width=0.95\linewidth]{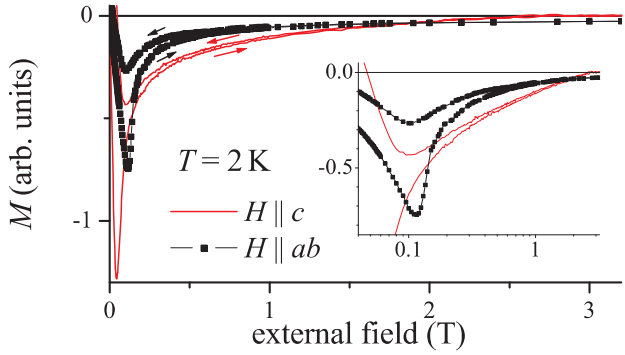}
\caption{Magnetization $M$ vs field $\mu_{\circ} H$ at $2\,{\rm
K}$ on a MgB$_2$ single crystal with $H\|c$ and $H\|ab$.}
\label{nFig1}
\end{figure}

The magnetization of the single crystal was measured as a function
of $H$; as can be seen in Fig.\ \ref{nFig1}, the irreversibility
is very low above about $0.15\,{\rm T}$. Larger irreversibility in
lower fields may be due to geometrical barriers, which is why we
did not attempt to directly extract $H_{c1}$. The curves shown are
not corrected for demagnetizing effects (the exact demagnetizing
factor is difficult to estimate due to the sample shape), but we
verified that any reasonable demagnetization correction does not
noticeably affect above $0.2\,{\rm T}$ the results discussed
below.

\begin{figure}[tb]
\includegraphics[width=0.95\linewidth]{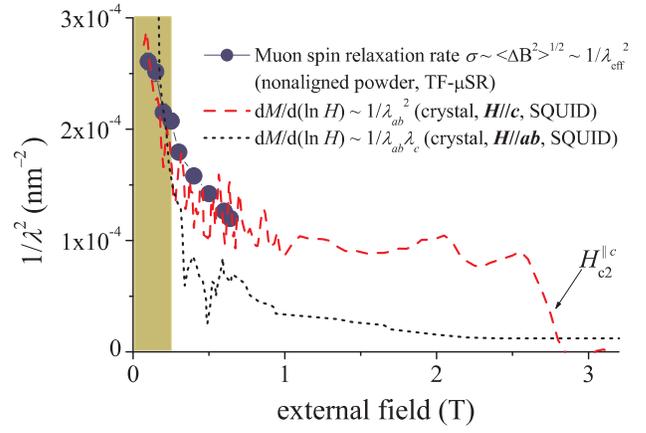}
\caption{Comparison of $1/\lambda^2$ vs $H$ obtained from ${\rm
d}M/{\rm d}\ln H$ of Fig.\ \ref{nFig1}, and from the muon spin
depolarization rate measured on unaligned powder (circles). The
shaded box indicates fields close to or lower than $H_{\rm{c1}}$
(see text).} \label{nFig2}
\end{figure}

Within the London model of a standard superconductor, the
magnetization is proportional to the logarithm of the applied
field, not too close to either $H_{c1}$ or $H_{c2}$. Keeping to an
analysis within the London approach, but dropping the requirement
of a constant penetration depth $\lambda$, we have $1/\lambda^2
\propto {\rm d}M_{\rm rev}/{\rm d}\ln H$. The so obtained
$1/\lambda^2$ is plotted in Fig.\ \ref{nFig2}. To avoid
overloading the graph, only curves assuming $M_{\rm rev} \! = \!
(M_{H \uparrow} \! + M_{H \downarrow})/2$ are shown; except in the
lowest $H$ using instead $M_{H \uparrow}$ or $M_{H \downarrow}$
leads to very similar results. The curves for both $H$ directions
were normalized by the same constant factor. The shaded box
indicates the low $H$ region, where we are uncertain about the
obtained penetration depth because of i) irreversibility, ii)
demagnetizing effects, and iii) deviations from the London model
due to the vicinity of $H_{c1}$ (see below).

Also plotted in Fig.\ \ref{nFig2} is $1/\lambda^2$ obtained from
the muon spin depolarization rate $\sigma$ measured on randomly
aligned powder at the same temperature. The depolarization rate
$\sigma$ is a measure of the average variation of the internal
field within a superconductor, and in the mixed state (again not
too close to one of the critical fields) is directly proportional
to $1/\lambda^2$, since $\lambda$ is the fundamental length scale
of the variation of the field in a superconductor (cf.\ Ref.\
\onlinecite{DiCastro03}). An issue to be aware of when
deducing $\lambda$ in this way 
is the possible influence of pinning, which can lead to an
extrinsic increase in $\sigma$. In a previous $\mu$SR experiment
on MgB$_2$, the whole $H$ dependence of $\sigma$ was indeed
ascribed to pinning.\cite{Niedermayer02} To check for the possible
influence of pinning on $\sigma$ (as opposed to the
magnetization), we performed time-dependent measurements of
$\sigma$ in several fields: After reaching $2\,{\rm K}$ (field
cooled), statistics was gathered for $10\,{\mathrm{min}}$, then
stopped and restarted (repeated $5$ to $10$ times). Except for
$0.1\,{\rm T}$, changes of $\sigma$ with time are well below error
bars, and no clear trend discernible. This suggests that for
higher $H$ even at $2\,{\rm K}$, pinning is not influencing
$\sigma$ much, and the $H$ dependence of $\sigma$ indeed
intrinsic. That we observed a very similar $H$ dependence of
$\sigma$ in samples from two sources synthesized slightly
different supports this; an intrinsic $\sigma (H)$ dependence was
also proposed in Ref.\ \onlinecite{Serventi04}.

Concerning the ${\mathrm{d}}M/{\mathrm{d}}(\ln H)$ curves, it may
be argued that the low field behavior is not unexpected even for a
normal superconductor, since in the limit $H \! \rightarrow
H_{c1}$ it is expected \cite{Fetter69}
${\mathrm{d}}M/{\mathrm{d}}(\ln H) \propto H/(H\!-\!H_{c1})$.
However, ${\mathrm{d}}M/{\mathrm{d}}(\ln H)$ should reach values
close to the normal London slope rather quickly (within $2\!-\!3\,
H_{c1}$) and the variation presented in Fig.\ \ref{nFig2} is
spread over a considerably larger field range. Furthermore, the
influence of the vicinity of $H_{c1}$ on $\sigma$ is opposite. The
close similarity of the $H$ dependence of the penetration depth
obtained from rather different quantities (bulk magnetization from
SQUID and internal field variation from $\mu$SR) strongly suggests
that all curves in Fig.\ \ref{nFig2} indeed show the $\lambda (H)$
dependence, outside of the shaded box indicating vicinity of
$H_{c1}$. A similar strong depression of $1/\lambda^2$ with $H$
was deduced previously (for $H\|c$) from an analysis of SANS form
factors.\cite{Cubitt03b} The SQUID curves additionally indicate
that above $10\,{\rm kOe}$ $1/\lambda^2$ no longer varies
strongly.

The SQUID measurements in the two field configurations also yield
the anisotropy of $\lambda$. For $H\|c$ the screening currents
flow within the $ab$ plane, giving $1/\lambda_{ab}^2$. For $H\|ab$
the currents flow also perpendicular to the planes, giving
$1/(\lambda_{ab} \lambda_c)$. The ratio of the ${\rm d}M/{\rm
d}\ln H$ curves for the two field configurations thus corresponds
to $\gamma_{\lambda}$. Considering the curves in Fig.\
\ref{nFig2}, we can see that i) in low $H$ the anisotropy is very
small, ii) in high $H$ $\gamma_{\lambda}$ is of the order of about
$6$ or even $7$, and iii) the variation with $H$ of
$\gamma_{\lambda}$ is most pronounced in low $H$. We stress the
fact that when considering the high field region alone the
standard London model with constant $\lambda$ and $\xi$ and a
constant common anisotropy $\gamma$ describes the data reasonably
well. This indicates that at low $T$ in high $H$ MgB$_2$ is close
to a ``standard superconductor'' with high anisotropy.
Corresponding to this is the absence of an unusual
$H_{c2}(\theta)$ dependence at low $T$, in contrast to the
situation closer to $T_{c}$ \cite{Golubov03,Angst03Rio,Rydh04} [of
course the $\gamma_{\lambda}$ analysis breaks down as $\mu_{\circ}
H \! \rightarrow \mu_{\circ} H_{c2}^{\|c}$ ($\simeq 2.8\,{\rm T}$
for this crystal)].

An alternative method to determine the penetration depth
anisotropy is to analyze the angular $\theta$ dependence of the
torque $\tau$ in fixed $H$. We previously used this method at much
higher $T$, finding also indications of an anisotropy increasing
with $H$.\cite{MgB2anisPRL02} However, thermal fluctuations and
additional intermixture of the two sets of bands by thermally
excited quasiparticles, complicate the analysis there. To provide
a direct comparison with the SQUID results and give a quantitative
estimate of $\gamma_{\lambda}$ we measured $\tau(\theta)$ at low
temperature.

\begin{figure}[tb]
\includegraphics[width=0.95\linewidth]{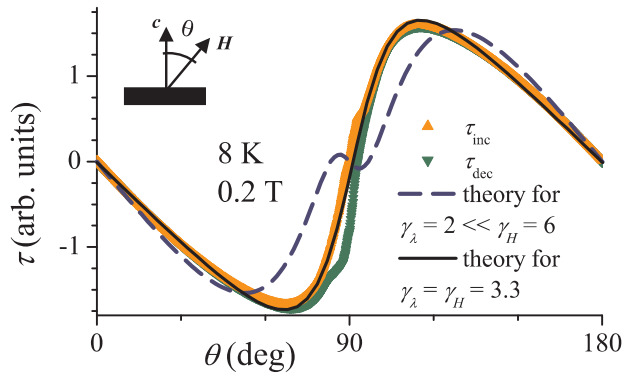}
\caption{Angle $\theta$ dependence of torque $\tau$ in $0.2\,{\rm
T}$ at $8\,{\rm K}$ (symbols). Dashed line: theoretical
description \cite{Kogan02b} assuming $\gamma_{\lambda} \ll
\gamma_H$; full line: description with $\gamma_{\lambda} =
\gamma_H$.} \label{nFig3}
\end{figure}

Due to increased irreversibility at lower $T$, it is important not
only to use a crystal with low pinning, but also employ the
``shaking technique'' developed by Willemin {\em et
al.}\cite{Willemin98} Since the magnetometer equipped with this
technique cannot reach $2\,{\rm K}$ measurements were conducted at
$8$, $11$ and $15\,{\rm K}$. These temperatures should be low
enough to avoid too strong an influence of thermal
fluctuations/excitations, as well as to probe the low temperature
limit of the calculated \cite{Kogan02,Miranovic03} anisotropies.

The data were analyzed with Eq.\ (18) of Ref.\
\onlinecite{Kogan02b}, allowing for a difference in the
anisotropies of the penetration depth and the coherence length.
Such a difference was, however, not found in any of the curves
analyzed, and a sign reversal of the torque, a key prediction of
Ref.\ \onlinecite{Kogan02b} for $\gamma_{\lambda} \ll \gamma_H$,
was never observed [for an example see Fig.\ \ref{nFig3}]. A
preliminary report on this issue is given in Ref.\
\onlinecite{Angst03Rio}. The best descriptions with Eq.\ (18) of
Ref.\ \onlinecite{Kogan02b} of the data were rather achieved for
$\gamma_{\lambda} \approx \gamma_H$. The large number of
parameters involved and the numerical condition of the fit formula
result in extended error bars though, so that {\em small}
differences between $\gamma_{\lambda}$ and $\gamma_H$ cannot be
completely ruled out (but large differences\cite{Kogan02b} as
calculated with $H$ independent anisotropies {\em can}).

\begin{figure}[tb]
\includegraphics[width=0.95\linewidth]{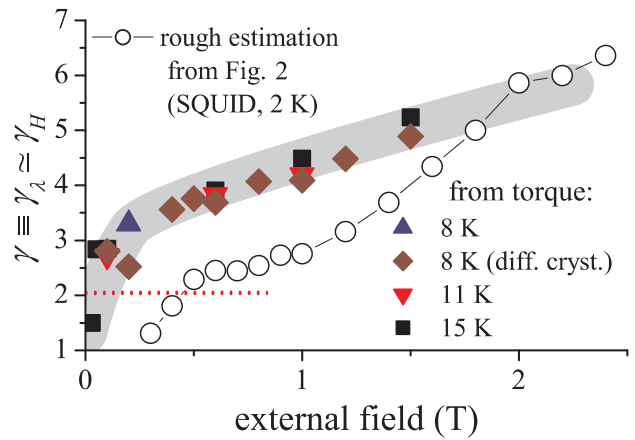}
\caption{Anisotropy determined from an analysis of the torque data
(Fig.\ \ref{nFig3}) with Eq.\ (18) of Ref.\ \onlinecite{Kogan02b},
at various low temperatures, as a function of field $H$. Also
shown is the anisotropy determined from Fig.\ \ref{nFig2}
($\circ$) and the anisotropy very close to $T_c$ (dotted line, see
text).} \label{nFig4}
\end{figure}

The resulting field dependence of the anisotropy $\gamma \equiv
\gamma_{\lambda} \approx \gamma_H$ is shown in Fig.\ \ref{nFig4}.
The anisotropy is monotonically increasing with increasing field,
up to $1.5\,{\rm T}$ (the maximum attainable by the magnetometer
used). In the lowest fields, this rise seems to be much steeper
than above $0.5\,{\rm T}$. However, as a cautionary note, even
with ``shaking'', the irreversibility cannot be said to be
negligible below $0.1\,{\rm T}$ and it should also be kept in mind
that we are approaching $H_{c1}$ (see Fig.\ \ref{nFig1}).

A report by another group of torque vs angle measurements
performed at $10\,{\rm K}$ claimed a field-independent anisotropy
of the order of $\gamma \simeq 4.3$.\cite{Takahashi02} However,
analyzing the same data, a different conclusion of an anisotropy
that {\em does} increase with $H$, in not too large fields, may
also be reached.\cite{CommentTakahashi} The results of Ref.\
\onlinecite{Takahashi02}, as well as Ref.\
\onlinecite{Zehetmayer02} (not finding a field-dependence as well)
may be reconciled with the ones of Ref.\
\onlinecite{MgB2anisPRL02} and the present results by assuming a
tendency of $\gamma$ to saturate in higher fields.

For comparison, a rough estimate of $\gamma (H)$ from SQUID
magnetometry in fields along the principal axes [Fig.\
\ref{nFig2}] is plotted in Fig.\ \ref{nFig4} as well.
Qualitatively ($\gamma (H)$ being an increasing function) this is
consistent with the torque results. We attribute the numerical
discrepancy to the large scattering as visible in Fig.\
\ref{nFig2} and the corresponding uncertainty in the estimation of
$\gamma$. The low $T$ behavior of $\gamma (H)$ is in strong
contrast to the one very close to $T_c$, where between $H_{c1}$
and $H_{c2}$ $\gamma \simeq 2$ is constant,\cite{note_Kohout} as
indicated by the dotted line in Fig.\ \ref{nFig4}.

A field dependent anisotropy at low $T$ ($2\,{\rm K}$) had been
deduced based on different experiments as well. Bouquet {\em et
al.}\cite{Bouquet02} reported a $H$ dependent effective anisotropy
based on specific heat measurements. Since these are sensitive
mainly to the coherence length, the experiment suggests the
anisotropy $\gamma_H$ to be $H$ dependent. Cubitt {\em et
al.}\cite{Cubitt03b} observed the anisotropy of the vortex lattice
$\gamma_{VL}$ to increase strongly with increasing $H$, from less
than $1.5$ in $0.2\,{\rm T}$ to about $3.8$ in $0.5\,{\rm T}$,
$\gamma_{VL}(H)$ being more steep in higher $H$. Keeping to the
London model, the anisotropy of the vortex lattice should be equal
to the penetration depth anisotropy
$\gamma_{\lambda}$.\cite{Campbell88} Our results extend to higher
$H$ and agree qualitatively with those of Cubitt {\em et al.}, but
we do not find a particularly steep $\gamma(H)$ around $0.5\,{\rm
T}$, but rather a slower field dependence. Very recently, Lyard
{\em et al.}\cite{Lyard04} proposed a similar $H$ dependence of
anisotropies, based on a London analysis of magnetization data
measured at much higher $T$.

The strong field dependence at low $T$ of all anisotropies
obtained from the measurements presented here, as well as by other
groups, are readily explained qualitatively by a faster
suppression with $H$ of superconductivity in the more isotropic
$\pi$ bands, increasing the relative contribution of the highly
anisotropic $\sigma$ bands. Such a suppression, consistent with
the overall decrease of $1/\lambda ^2$ (Fig.\ \ref{nFig2}), was
also observed, e.g., by spectroscopic means,\cite{Gonnelli02} and
is not unexpected due to the smaller gap in the $\pi$ bands.
Corresponding larger vortex cores and a lower ``$H_{c2}^{\pi}$''
have been conjectured from STS and specific heat
measurements.\cite{Eskildsen02,Bouquet02} If $\pi$ and $\sigma$
bands were independent, a real upper critical field $\mu_{\circ}
H_{c2}^{\pi} \approx 0.5\,{\rm T}$ would mark the destruction of
superconductivity in the $\pi$ bands due to vortex core overlap.
Since, however, the bands are coupled together even at zero $T$,
``$H_{c2}^{\pi}$'' degenerates into a broad crossover (completely
blurred for $T \! \rightarrow T_c$). Our results indicate that
this crossover region is very broad, extending down to almost zero
field. In high $H$ superconductivity in the $\pi$ bands is still
induced from the $\sigma$ bands likely up to the bulk $H_{c2}$,
but with a much depressed order parameter. It should be noted that
within this picture, there would in principle be two different
coherence lengths to consider,\cite{Serventi04} and that in the
$H$ region of interest, the vortex cores in the $\pi$ band overlap
\cite{Eskildsen02} enough to seriously question the applicability
of a London analysis. This may explain the remaining discrepancies
between the anisotropies obtained from different measurements and
calls for further theoretical work, although in a qualitative way
the London analysis works out remarkably well, particularly in
high $H$ at low $T$, in terms of a ``standard anisotropic''
$\sigma$ band only superconductor.

\label{conc} In conclusion, $\mu$SR and magnetization data show
the ``superfluid density'' $1/\lambda^2$ in MgB$_2$ at $2\,{\rm
K}$ to strongly decrease with increasing field below about
$1\,{\rm T}$. In parallel, the penetration depth anisotropy
increases, and is not drastically smaller than the coherence
length anisotropy (in the same field). This behavior is due to the
fast suppression of the contribution to superconductivity of the
more isotropic $\pi$ bands with weaker superconductivity.

\label{ack} Work partly performed at the Swiss Muon Source
(S$\mu$S) at the Paul Scherrer Institute (Villigen, Switzerland).
We thank D.~Herlach and A.~Amato for technical assistance during
the $\mu$SR experiments and V.~G. Kogan and A.~Gurevich for
discussions. This work was supported by the Swiss National Science
Foundation (SNSF) and by the NCCR program MaNEP sponsored by the
SNSF. Ames Lab is operated for the U.S. Department of Energy by
Iowa State Univ. under Contract W-7405-Eng-82. The work at Ames
was supported by the Director of Energy Research, Office of Basic
Energy Sciences.

\newcommand{\noopsort}[1]{} \newcommand{\printfirst}[2]{#1}
  \newcommand{\singleletter}[1]{#1} \newcommand{\switchargs}[2]{#2#1}


\begin{thebibliography}{10}
\bibitem{note_twobandanis}
We should note at this point that, due to the two-band nature of
{MgB$_2$}, in
  any case we are dealing with effective length scales, derived from the
  natural length scales within the $\sigma$ and the $\pi$ bands according to
  the relative contributions of the two sets of bands. An analysis of $\mu$SR
  data more directly related to different intrinsic $\sigma$ and $\pi$ length
  scales was recently proposed in Ref.\ \onlinecite{Serventi04}.
\end{thebibliography}
\end{document}